\documentclass[10pt,letterpaper]{article}

\usepackage{cogsci}

\cogscifinalcopy 

\usepackage{pslatex}
\usepackage{apacite}
\usepackage{graphicx}
\usepackage{float} 

\title{Institutional Preferences in the Laboratory}
 
\author{{\large \bf Qiankun Zhong (zhong@mpib-berlin.mpg.de)} \\
  Center for Humans and Machines, Max Planck Institute for Human Development \\
  Berlin 14195 Germany
   \AND {\large \bf Nori Jacoby ( kj338@cornell.edu)} \\
  Department of Psychology, Cornell University \\
  Ithaca, NY 14853 USA \\
  Max Planck Institute for Empirical Aesthetics\\
  Frankfurt 60322 Germany
   \AND {\large \bf Ofer Tchernichovski (otcherni@hunter.cuny.edu)} \\
  Department of Psychology, Hunter College, CUNY\\
  New York, NY 10065 USA 
  \AND {\large \bf Seth Frey (sethfrey@ucdavis.edu)} \\
  Department of Communication, University of California, Davis \\
  Davis, CA 53706 USA}

\begin{document}

\maketitle

\begin{abstract}
Getting a group to adopt cooperative norms is an enduring challenge. But in real-world settings, individuals don't just passively accept static environments, they act both within and upon the social systems that structure their interactions. Should we expect the dynamism of player-driven changes to the "rules of the game" to hinder cooperation --- because of the substantial added complexity --- or help it, as prosocial agents tweak their environment toward non-zero-sum games? We introduce a laboratory setting to test whether groups can guide themselves to cooperative outcomes by manipulating the environmental parameters that shape their emergent cooperation process. We test for cooperation in a set of economic games that impose different social dilemmas. These games vary independently in the institutional features of stability, efficiency, and fairness. By offering agency over behavior along with second-order agency over the rules of the game, we understand emergent cooperation in naturalistic settings in which the rules of the game are themselves dynamic and subject to choice. The literature on transfer learning in games suggests that interactions between features are important and might aid or hinder the transfer of cooperative learning to new settings.

\textbf{Keywords:} 
collective intelligence, cooperation, institutional evolution
\end{abstract}

\section{Introduction}

One of the main purposes of institutions — "the humanly devised constraints that structure political, economic, and social interaction" \cite{North1991} — is to formalize collaboration and solve collective problems. But in practice, we see a startling diversity of solutions for any given problem \cite{Ostrom2005, Shivakoti1993}. What type of institutions should we choose to solve the collective problems? Given the diversity of options, do people collectively prefer one over others? Do their choices or preferences entail some underlying cultural values or principles? 

Behavioral research on cooperation overwhelmingly attends to static conditions, disregarding the fact that a distinguishing feature of humans is their command over their environment. From stigmergy to mechanism design, we engage in second-order manipulation of our social environments, often preventing cooperation failures not by requiring prosocial behavior but by creating win-win environments in which the natural outcome is cooperation. John Rawls opened questions of social design to inquiry by positing the veil of ignorance, a scenario in which members of a theoretical society design its roles blind to the role that each member will fill \cite{rawls1971theory}. He uses this set up to offer a just society as one that a person would accept in ignorance of their own eventual position in it. Rawls' thought experiments were brought to the laboratory by Tetlock, who offered fictional narrative scenarios to investigate how people trade off between social values such as fairness and efficiency \cite{tetlock1986value,mitchell2009disentangling}. However, in his designs participants were never embedded in the incentive systems that they expressed preferences for, and thus had no opportunity to close the loop between their first-order experiences of the results of their second-order institutional preferences. Elaborating the theories of Rawls and others, Binmore addresses institutional characteristics directly, proposing explicitly that social development will reflect preferences primarily for stability, then for efficiency, and third for fairness, all else being equal \cite{binmore2005natural}. Frey and Atkisson formalized a scenario of game change in a series of simulations over a large game space, representing dynamics through institution space to understand the importance of fairness \cite{Frey2020}. In their simulations, strictly selfish agents are placed into a random game and iteratively permitted to select neighboring games that are preferred in terms of stability over predictability over efficiency, with no social preferences. They find that the attractors of agents' search trajectories are more likely to be fair than other games in the space, despite an absence of social preferences, because the win-win games were a substantial fraction of the otherwise competitive games they converged upon.

Outside of formal environments, economists and anthropologists have taken their own interest in the role of individual preferences and choices in institutional evolution. Economists recognized that appropriate institutional arrangements depend on the initial conditions of a society or an organization \cite{djankov2003new} and sometimes the path toward an optimal institution might be sequential and the costs to implement institutional functions are unbalanced. The challenge is to decide on which path to go for when we settle for a second-best institution and which constraints to remove to achieve desired outcomes. Anthropologists point out that individuals' collective preferences can serve as an important selective force guiding  institutional evolution toward group-beneficial norms and institutions. Cultural group selection theory proposed a pathway to overcome the lock-in situation and shift towards another equilibrium in the long run\cite{baum2007cultural}. Cultural group selection theory explains how individually costly cultural traits can evolve due to the competitive advantage they bestow at the group level \cite{henrich2004cultural, boyd2010transmission}. For example, executing third-party punishment can be costly for individuals but provides a public good to their group, which leads to better group performance in the long run \cite{Richerson2016}. At the same time, individuals' preferences emerge from their interactions with others and the environment \cite{bowles1998endogenous,fouka2020backlash,zhong2022institutional}. Wildavsky argues that individuals form diverse preferences by using their social relationships to interpret their environment, with culture as a more powerful construct than alternatives like heuristics, schemas, or ideologies \cite{wildavsky1987choosing}. 

Previous studies on institutional development focus either on the external pressure of between-group competition or the internal mechanics of within-group collaboration. Few studies have connected these levels of analysis to explain how within-group behavior can shift between-group  dynamics. Individuals’ preferences and choices can be an important selective force guiding the development of efficient institutions (e.g., if people vote for institutional structures with their feet) \cite{Boyd2009}. 

Empirically investigating individuals' institution-level preferences is difficult. The first challenge is to unbundle the broad cluster of institutions and learn more about the relative importance of institutional features  \cite{acemoglu2005unbundling}. The second challenge is to find valid proxies for two or more sets of institutions \cite{jancsary2017toward}. The third challenge is measuring collective preferences and estimating their influence on institutional evolution. This paper is an attempt to address these challenges through laboratory experiments. We conducted a series of experiments in which participants chose between "toy" institutions (simple economic games) to reveal what qualities they preferred in their institutions. By observing which games people prefer, we can infer participants’ internal tradeoffs between features like stability, efficiency, and fairness and, from there, model how those preferences drive bottom-up institutional change. We ask:





RQ1: How are cooperative behaviors likely to evolve within different institutional environments?

RQ2: Are individual preferences structured in a way that following them naturally leads to the emergence of optimal institutions?

\section{Method}

In this study, we conducted an online experiment in which participants played two simple games and decided which game they preferred. After selecting one of the two games, they had the opportunity to play the game they choose. We then estimated how cooperation emerged within the games, and which types of institutional features individuals selected.

\subsection{Game Design}
Building on previous simulations on institutional preference \cite{Frey2020}, we design a space of 8 simple economic games in which the game-level qualities of stability, efficiency, and fairness can be varied independently. Within this space, a game is fair when the summed payoffs of all outcomes are equal for both players, efficient when the sum of pay-offs is high, and stable when there is only one unique Nash equilibrium. To create forced choices between the three features, we then constructed 12 pairs of games for players to compare. The games and the comparison pairs are illustrated in Figure 1.

\begin{figure}[ht]
\begin{center}
\includegraphics[scale=0.42]{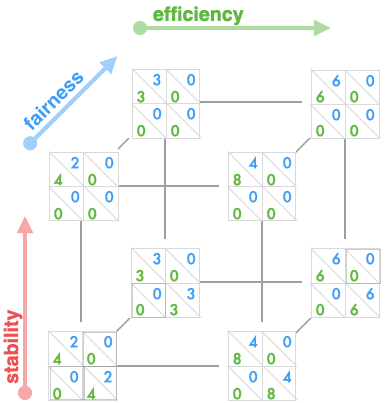}
\end{center}
\caption{Game Design} 
\label{game-design}
\end{figure}

Each game has two players. The first player will choose between the top and bottom rows and the second player will choose between the left and right columns. Each participant is equally likely to play from the perspective of the first or second player to ensure that participants understand the game from both perspectives. As some of the games are asymmetric, to disentangle the effect of row/column and first/second players, we also transpose the game matrix to make sure that as a row player, the player can access both Player 1's and Player 2's payoffs. To avoid biases introduced by the spatial arrangement of the game, we transform the games in three ways, including switching rows, switching columns, and transposing.
\subsection{Sampling and game structure}
The experiment used a Dallinger-based online experiment platform \textit{PsyNet} \cite{harrison2020gibbs} to recruit players and assign them randomly into various game conditions. All participants were recruited through Prolific, compensated at a rate of \$9 per hour, and took part in the experiment after providing informed consent in accordance with an approved protocol (Max Planck Ethics Council \#202142). This project is structured in a way that each pair of two games at one of their 8 transformations is defined as an experiment condition and the participants will be assigned randomly to play a pair of games. 

To ease recruitment burdens, and leverage the benefits of our experimental platform, we design the experiment for an asynchronous gameplay approach to ensure a large and balanced sample for each condition. Participants enter a “virtual room” where all game boards are the same type and some have already been played. They are then “seated,” meaning they are virtually assigned to one of these ongoing games. This way, they can take on the role of Player 1 if they are the first to join a table or Player 2 if they are joining an ongoing game. In the latter case, we can then reveal the entire gameplay to them. After completing a round, they are reassigned to a new “table” (and potentially assigned to a different role). This setup allows participants to experience playing with others without requiring everyone to be present simultaneously.

To incentivize participants, we also offer a bonus relative to players' payoff in each game. For Player 2s, we provide spontaneous results and a bonus when Player 2 makes the choice. For Player 1, we do not know the results in real-time. To provide a spontaneous bonus, we estimate the probability of Player 2’s choice based on the data collected previously and pay the participant the estimated bonus.



\subsection{Procedure}
\begin{figure}[ht]
\begin{center}
\includegraphics[scale=0.4]{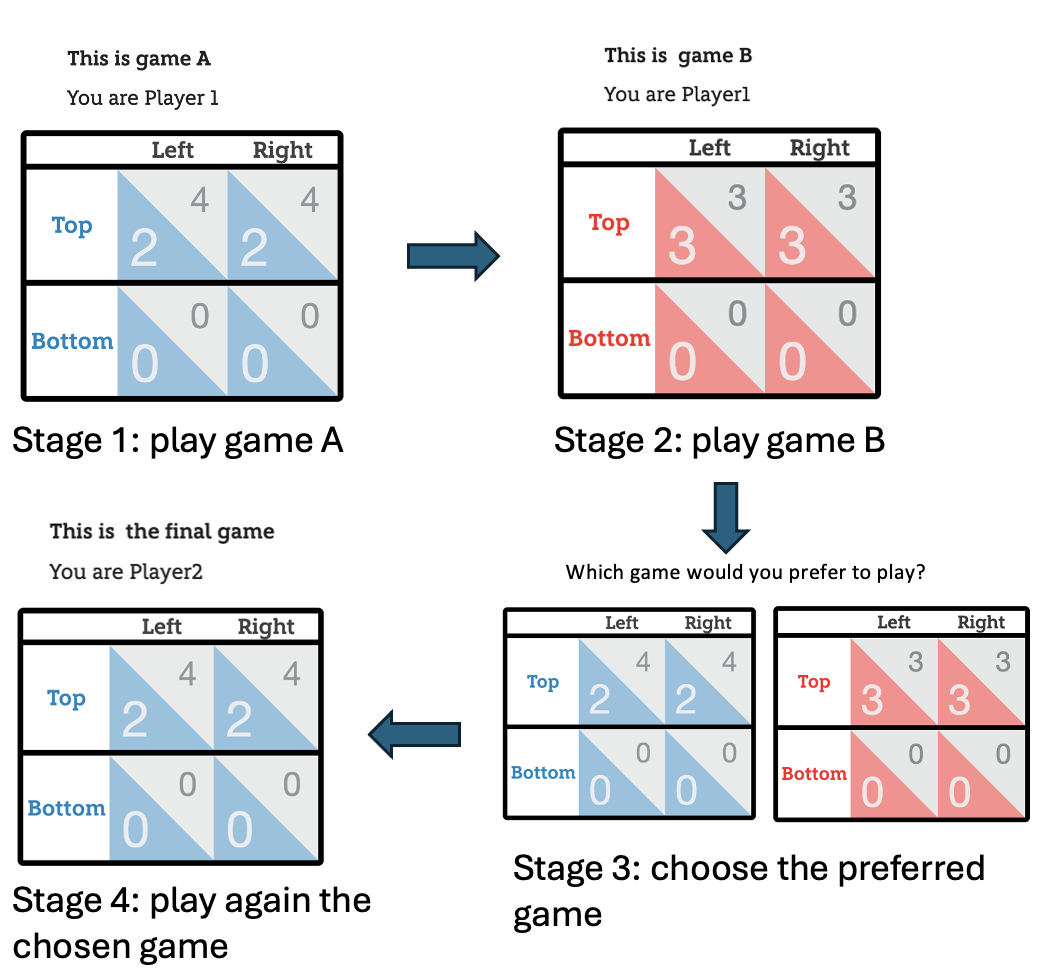}
\end{center}
\caption{Game procedure} 
\label{procedure}
\end{figure}

After entering the game, each participant went through a tutorial to learn how to read their payoffs from the game table; then they were asked to complete a comprehension quiz. After the tutorial and quiz, they entered the actual gameplay (see Figure \ref{procedure}). In Stage 1, participants repeated the first game six times. Each time, the participant had an equal chance to be Player 1 or 2, and the game was displayed in one of the eight transformations. In Stage 2, the participant played the second game with a different color similarly for 6 times. In Stage 3, the participant was presented with a version of both games to select the game they prefer to play. After they selected the game, they then played the game they chose 6 more times as Stage 4. After they finished the gameplay, they were asked to fill in demographic information and finish a closing comprehenion quiz similar to the initial comprehension quiz.

\subsection{Analysis}
We collected data on the players' choices within each game and their preferences between the two games. To analyze the within-game dynamics, we define cooperation by any non-zero outcomes and calculate the cooperation rate of each game. To analyze the between-game dynamics, we estimate the proportion that each game is selected. 

\section{Results}
\begin{figure}[ht]
\begin{center}
\includegraphics[scale=0.45]{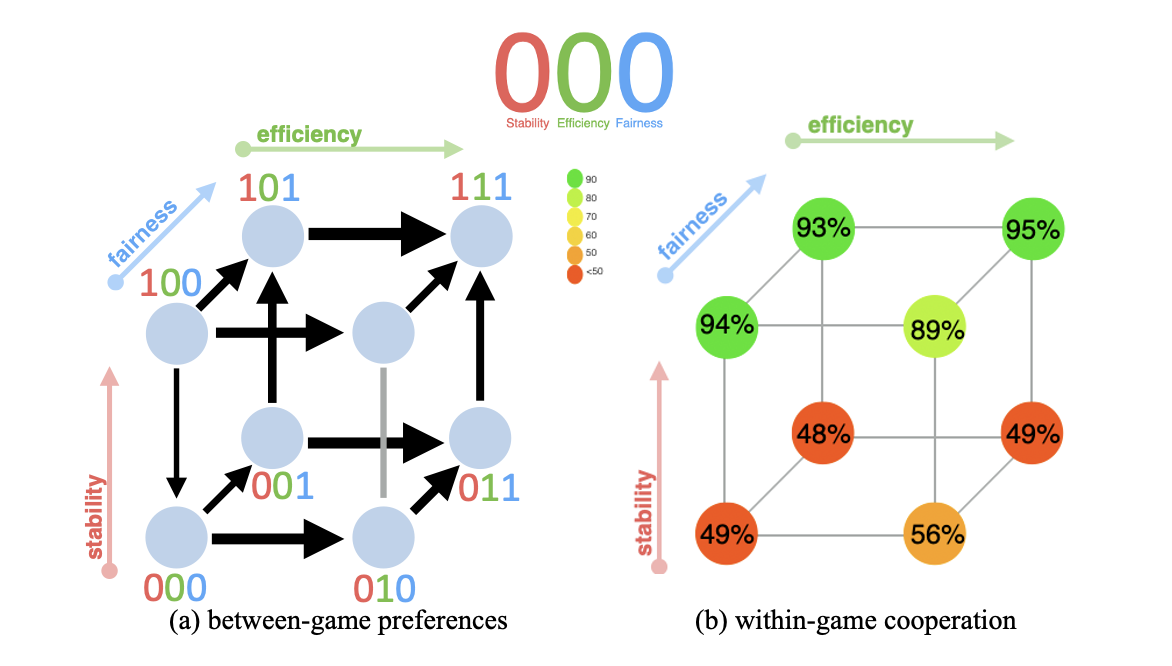}
\end{center}
\caption{Comparing between-game preference and within-game cooperation} 
\label{resultall}
\end{figure}

We recruited a total of 310 participants with 3951 game trials. After removing 409 trials playing against the initializing seeds, we conclude with 3542 trials across 8 different games. To structure our analysis, we categorize games with none of the institutional features (000) as the first layer of games, games with one institutional feature (100, 010,001) as the second layer, with two institutional features (110, 101, 011) as the third layer, and with three institutional features (111) as the fourth layer (see Figure \ref{resultall}). Because all game features are positive, games in higher layers are expected to be preferred over games in lower layers. We then analyze the collective preference between games in different layers and the within-game cooperation rate. We also bootstrap the data to get the $ 95 \%$ confidence interval for statistical comparison.
\subsection{Between-game Results}
\begin{figure}[ht]
\begin{center}
\includegraphics[scale=0.5]{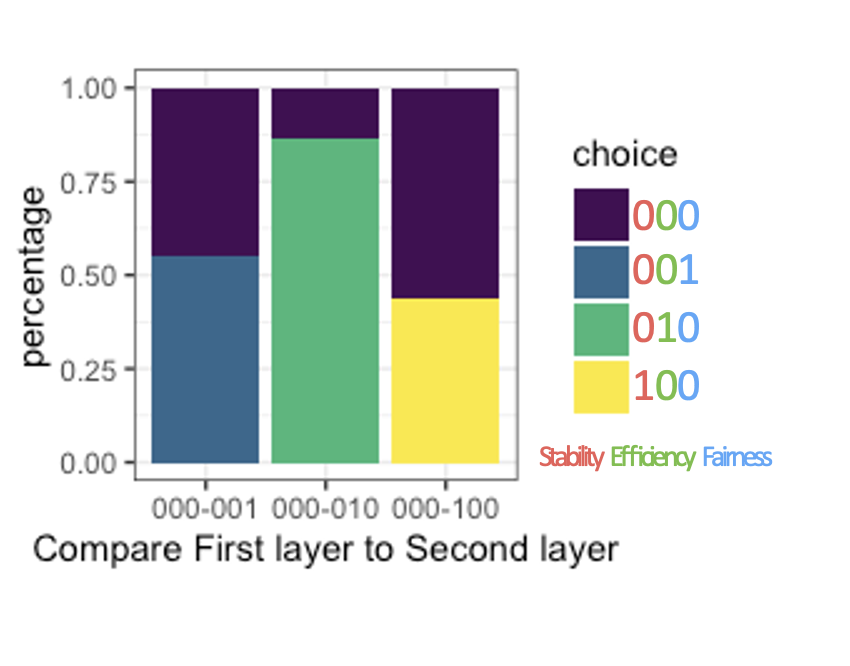}
\end{center}
\caption{Between Game Comparison between First and Second Layer} 
\label{between1}
\end{figure}

After removing 9 participants who had encountered technical errors before comparison, we ended up with 301 participants who chose between the two games they were presented with. We first look into the first feature people prefer to have while in a null state of no stability, efficiency, or fairness (000). Figure \ref{between1} demonstrates the three comparisons where, except for efficiency, participants don't have a strong preference for the first layer of games.
We then look into the preference for the institutional styles conditional on the existence of stability (100), efficiency (010), or fairness (001).
\begin{figure}[ht]
\begin{center}
\includegraphics[scale=0.35]{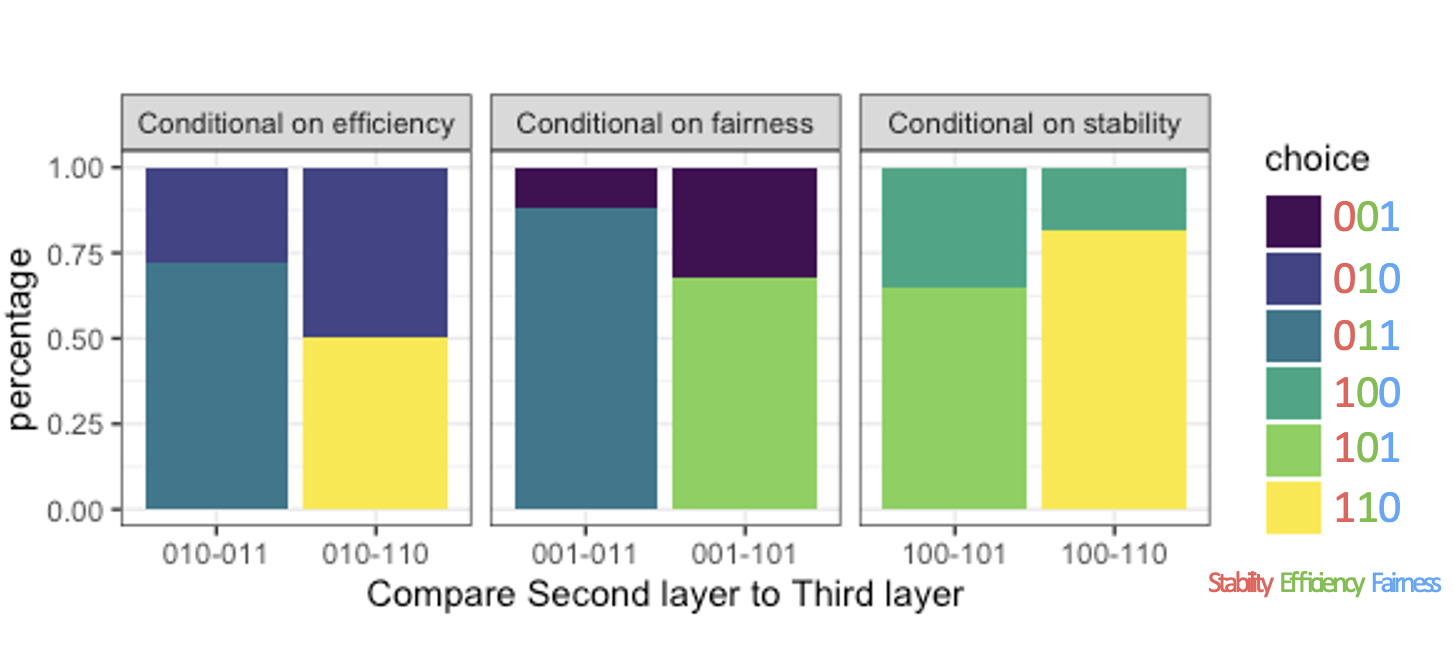}
\end{center}
\caption{Between Game Comparison between Second and Third Layer} 
\label{between2}
\end{figure}
We finally compare the third layer to the ultimate game with all three features (see Figure \ref{between3}). 
\begin{figure}[ht]
\begin{center}
\includegraphics[scale=0.48]{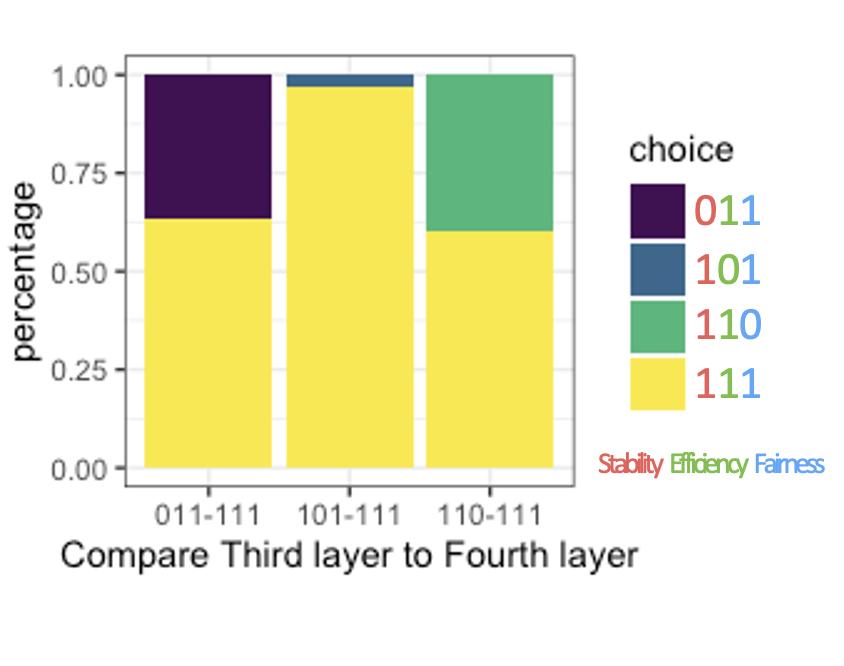}
\end{center}
\caption{Between-Game Comparison between the Second and Third Layers} 
\label{between3}
\end{figure}
\subsubsection{Inconsistent gradients}
There is a clear stepwise increase when we go from layer 1 to layer 3 in order of having stability first and efficiency next (000 – 100 – 110). Compared to the null state 000, implementing stability first is not strongly preferred ($p_{100} = 0.43$, $CI=[0.22,0.65]$), then after implementing stability (100), the next step of implementing efficiency is a clear jump of preference($p_{110} = 0.81$, $ CI=[0.64,0.95]$). But with implementing efficiency first and stability next ( 000 — 010 — 110), we see the preference reaches a peak with $p_{010} = 0.86$, $CI=[0.73,1]$  and then plateaus with $p_{110} =  0.5$, $CI=[0.29,0.71]$.

Similar patterns appear in the paths to stability from fairness and efficiency. We can compare the path of fairness ($p_{001} = 0.55$, $CI=[0.38,0.72]$) — efficiency ($p_{011} = 0.88$, $CI=[0.76,1]$) — stability ($p_{111} = 0.63$, $CI=[0.45,0.82]$) and efficiency ($p_{010} = 0.86$, $CI=[0.73,1]$) — fairness ($p_{011} = 0.55$, $CI=[0.38,0.72]$) — stability ($p_{111} = 0.63$, $CI=[0.45,0.82]$) and see a clear asymmetric path of preference due to the main effects of efficiency and the interaction effect of stability.

We last compare fairness—stability (000—001—101) and stability—fairness (000—100—101). The result suggests that these two paths are relatively more balanced. Implementing stability first leads to a steeper jump in preference ($p_{100} = 0.43$, $CI=[0.22,0.65]$, $p_{101} = 0.65$, $CI=[0.43,0.83]$) compared to implementing fairness first ($p_{001} = 0.55$, $ CI=[0.38,0.72]$, $p_{101} = 0.67$, $CI=[0.52,0.84]$). 

\subsubsection{Main Effects}

Overall, there is a strong preference for efficiency across all scenarios. In Figure 1, at null state ($000$), the preference for a game with efficiency only ($p_{010} = 0.86$, $CI=[0.73,1]$) is much higher than a game with fairness ($p_{001} = 0.55$, $CI=[0.38,0.72]$ ) or with stability ($p_{100} = 0.43$, $CI=[0.22,0.65]$) only.
As shown in Figure \ref{between1}, conditioning on having fairness in the game ($001$), the game with efficiency has a much higher appeal ($p_{011} = 088$, $ CI=[0.76,1]$) than the games with fairness ($p_{101} = 0.67$, $CI=[0.52,0.84]$). Similarly, conditioning on having stability ($100$), the game with efficiency ($p_{110} = 0.82$, $CI=[0.64,0.95]$) is much more preferred than the game with fairness ($p_{101} = 0.65$, $CI=[0.43,0.83]$).

\subsubsection{Interaction Effects}

Fairness  ($p_{001} = 0.55$, $CI=[0.38,0.72]$) is not strongly preferred in the comparison between the first and second layers and gains much more preference conditioning on the existence of other features. When a game already has efficiency (010), the preference for fairness ($p_{011} = 0.72$, $CI=[0.55,0.86]$) has increased significantly. 

Stability ($p_{100} = 0.43$, $CI=[0.22,0.65]$) is also not preferred when compared to the null state. When conditioning on efficiency, the preference ($p_{110} = 0.5$, $CI=[0.29,0.71]$) only increased marginally, yet the preference is stronger ($p_{101}= 0.67$, $CI=[0.52,0.84]$)  for stability when conditioning on fairness (001). 

\subsection{Within-game Results}
\begin{figure}[ht]
\begin{center}
\includegraphics[scale=0.4]{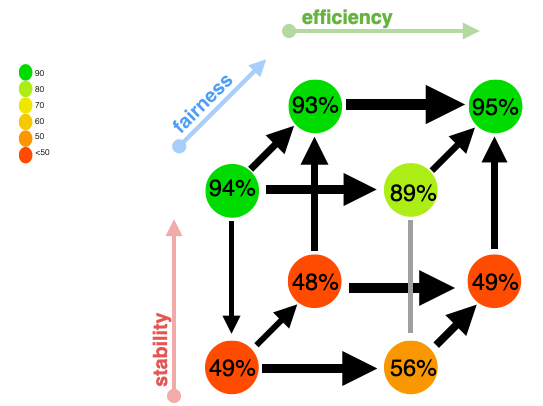}
\end{center}
\caption{Within Game Cooperation Rate} 
\label{within}
\end{figure}

To measure the cooperation rate, we remove all the Player 2 results against initializing seeds (human vs. random choices). We then estimated the cooperation rate for each game and listed the results in Figure \ref{within}.

When we compare cooperation rates between layers, it is clear from Figure \ref{within} that 
higher levels of games don't necessarily lead to higher cooperation rates. For example, Game 010 has a higher cooperation rate ($C_{010} = 0.56$, $CI=[0.49,0.62]$) than Game 011 ($C_{011} = 0.49$, $CI=[0.43,0.56]$). However, higher layers do have a higher average across games than lower layers.  

We then examine the specific features that might contribute to the cooperation rate. Stability in general leads to a higher cooperation rate, comparing Game 000 and 100 ($C_{000} = 0.49$, $CI=[0.43,0.56]$, $C_{100} = 0.94$, $CI=[0.90,0.97]$), Game 010 and 110 ($C_{010} = 0.56$, $CI=[0.49,0.62]$ and $C_{110}=0.89$, $CI=[0.84,0.93]$), Game 001 and Game 101 ($C_{001} = 0.48$, $CI=[0.41,0.54]$, $C_{101}=0.93$, $CI=[0.90,0.96]$), Game 011 and 111 ($C_{011} = 0.49$, $CI=[0.43,0.56]$, $C_{111}=0.95$, $CI=[0.92,0.97]$). This effects driven by stability can be explained by the unique Nash Equilibrium design in the games with stability. In other words, there are no better alternatives for the players than cooperation.

Efficiency does not increase cooperation by clear margins, comparing Game 000 and 010 ($C_{000} = 0.49$, $CI=[0.43,0.56]$, $C_{010} = 0.56$, $CI=[0.49,0.62]$), Game 100 and 110 ($C_{100} = 0.94$, $CI=[0.90,0.97]$, $C_{110} = 0.89$, $CI=[0.84,0.93]$), Game 001 and 011 ($C_{001} = 0.48$, $CI=[0.41,0.54]$, $C_{011} = 0.49$, $CI=[0.43,0.56]$), Game 101 and 111 ($C_{101} = 0.93$, $CI=[0.90,0.96]$, $C_{111} = 0.95$, $CI=[0.92,0.97]$). 

Fairness only increases cooperation marginally conditional on both stability and efficiency (($C_{110} = 0.89$, $CI=[0.84,0.93]$, $C_{111} = 0.95$, $CI=[0.92,0.97]$). It does not increase the cooperative rate in other conditions, comparing Game 000 and 001 ($C_{000} = 0.49$, $CI=[0.43,0.56]$, $C_{001} = 0.48$, $CI=[0.41,0.54]$), Game 010 and Game 011  ($C_{010} = 0.56$, $CI=[0.49,0.62]$, $C_{011} = 0.49$, $CI=[0.43,0.55]$), and Game 100 and 101 ($C_{100} = 0.93$, $CI=[0.90,0.97]$, $C_{101} = 0.93$, $CI=[0.90,0.96]$).

\section{Discussion}
\subsection{Summary}
In this research, we constructed an online experiment that surveys collective institutional preference through simple economic games with variations in stability, efficiency, and fairness. We found that efficiency is preferred at all levels. Fairness is preferred only when efficiency is first implemented, while stability is desired only when both fairness and stability are preferred. 

This interaction effect between game features also leads to inconsistent paths towards optimal institutions, where implementing stability, efficiency, and fairness sequentially has a clear stairwise increased preference. Yet implementing efficiency first might lead to institutional lock-in where preferences for fairness and stability do not manifest. This sequential effect is important because only increased preference can drive cultural evolution toward an optimal institution when the selective force is strong in the macro environment. 

Notably, a higher cooperation rate within a game does not necessarily lead to higher preference for that game. Games with stability lead to a higher cooperative rate (e.g., $C_{100} = 0.94$), yet it is not preferred when compared to the null state game ($p_{100} = 0.43$). Conversely, games with the efficiency feature do not foster more cooperation (e.g., $C_{010} = 0.56$), but are preferred by participants regardless ($p_{010} = 0.86$). This indicates a behavioral bias towards efficiency even when participants may benefit less from games with that feature. This suggests that between-game preferences do not necessarily lead to cooperative norms or institutions. 
\subsection{Contribution}

This experiment demonstrates how individuals’ preferences for the features of a social system can favor and stabilize different norm sets, driving between-group evolution towards optimal forms of institutions beneficial for collaboration, equitable outcomes, or group efficiencies. The experiment results also address whether group-beneficial institutions are selected at the group level, at least in part, through individual choices. The current results do not support this theory, which calls for alternative design mechanisms of a robust system that encourages group-beneficial norms through personal choices.

This experiment provides empirical support for a mechanism of cultural group selection and has important implications for designing robust adaptive systems. The main ingredient for robust institutions is adaptive efficiency \cite{North1991, Bednar2018}. To date, studies that endogenize institutional change focus only on how the system responds to external disturbance or how individuals adapt to the system. Scholars are only beginning to understand how social-scale outcomes are affected by individuals’ preferences for certain institutional features, the interactions of features with each other, and the interactions of within- and between-game decisions. The experiment results suggest that evolutionary paths that are less likely to be locked in and less costly will be more likely to result in desired group benefits.

\subsection{Limitations and Future Work}
Among all the results, participants prefer games with the property of efficiency --- high total relative payoffs --- at all levels over the other two institutional features. This might coincide with real-world institutional goals, but the effect might also be amplified by how we implement efficiency in the game. In the current design, we activate the game property of efficiency by doubling all payoffs in the game matrix. In future designs, we will weaken the increase in efficiency to a multiplier of just 1.5x. 

So far, we have asked participants to indicate their preference for games in different layers. In other words, we are testing whether adding an institutional feature will result in a collective preference to switch to the new game. This can be considered a sanity check for whether people prefer more optimal games. At the same time, by comparing the new preferences in two paths (for example, conditioning on 001, comparing $p_{101}$ and $p_{011}$), we can deduce the preference between Game 101 and Game 011. In future experiments, we will expand the experiment by making people choose directly within a layer, such as comparing Game 101 and 011. This expansion will isolate the topological effects of the game from the effects of path dependency and allow us to construct a full picture of the institutional landscape. Last, to test whether game spaces can have different attractor basins depending on the starting game, will we will add a fourth orthogonal game feature, to create a 16-game game space organized along the corners of a four-dimensional hypercube.

As we pointed out in the results, the observed within-game cooperation rates are not consistent with between-game preferences: people in some cases preferred games with worse outcomes. One possible explanation is that participants did not have enough experience in each game tomake a clear, rational evaluation of the long-term benefit of a game. In future experiments, we may want to compare a short version of this game (6 repeated games) and a long version of this game (15--20 repeated games) to test whether individual behavioral bias can be shifted through long-term repeated interaction within a group. We will develop measures for system robustness and evolutionary stability to understand better the underlying logic of collective institutional preference.

In conclusion, work on preferences and decisions in social settings has generally assumed a static or exogenously changing social system, neglecting that people often have agency not only over their behavior within a game but over what game they are behaving within. This opens the possibility that people can bring a community to socially beneficial outcomes not just through within-game norms of behavior but by iteratively guiding institutional evolution to new games that foster cooperation more naturally. Using this mechanism requires designs in which the games are mutable. The benefit of research in this direction is that it opens up research questions to cognitive scientists. With this work, we can gain new insight into the decisions across levels of a social system, inquiring into the qualities that people seek in a social system, as well as the effects of giving communities agency over the conditions of their interactions. By introducing social situatedness into the dynamics of cooperation, we increase the validity and sophistication of experimental inquiry into individual contributors to social complexity.

\bibliographystyle{apacite}

\setlength{\bibleftmargin}{.125in}
\setlength{\bibindent}{-\bibleftmargin}

\bibliography{CogSci_Template}

\begin{thebibliography}{}

\bibitem [\protect \citeauthoryear {%
Acemoglu%
\ \BBA {} Johnson%
}{%
Acemoglu%
\ \BBA {} Johnson%
}{%
{\protect \APACyear {2005}}%
}]{%
acemoglu2005unbundling}
\APACinsertmetastar {%
acemoglu2005unbundling}%
\begin{APACrefauthors}%
Acemoglu, D.%
\BCBT {}\ \BBA {} Johnson, S.%
\end{APACrefauthors}%
\unskip\
\newblock
\APACrefYearMonthDay{2005}{}{}.
\newblock
{\BBOQ}\APACrefatitle {Unbundling institutions} {Unbundling institutions}.{\BBCQ}
\newblock
\APACjournalVolNumPages{Journal of political Economy}{113}{5}{949--995}.
\PrintBackRefs{\CurrentBib}

\bibitem [\protect \citeauthoryear {%
Baum%
}{%
Baum%
}{%
{\protect \APACyear {2007}}%
}]{%
baum2007cultural}
\APACinsertmetastar {%
baum2007cultural}%
\begin{APACrefauthors}%
Baum, J\BPBI A.%
\end{APACrefauthors}%
\unskip\
\newblock
\APACrefYearMonthDay{2007}{}{}.
\newblock
{\BBOQ}\APACrefatitle {Cultural group selection in organization studies} {Cultural group selection in organization studies}.{\BBCQ}
\newblock
\APACjournalVolNumPages{Organization Studies}{28}{1}{37--47}.
\PrintBackRefs{\CurrentBib}

\bibitem [\protect \citeauthoryear {%
Bednar%
\ \BBA {} Page%
}{%
Bednar%
\ \BBA {} Page%
}{%
{\protect \APACyear {2018}}%
}]{%
Bednar2018}
\APACinsertmetastar {%
Bednar2018}%
\begin{APACrefauthors}%
Bednar, J.%
\BCBT {}\ \BBA {} Page, S\BPBI E.%
\end{APACrefauthors}%
\unskip\
\newblock
\APACrefYearMonthDay{2018}{}{}.
\newblock
{\BBOQ}\APACrefatitle {When Order Affects Performance: Culture, Behavioral Spillovers, and Institutional Path Dependence} {When order affects performance: Culture, behavioral spillovers, and institutional path dependence}.{\BBCQ}
\newblock
\APACjournalVolNumPages{American Political Science Review}{112}{1}{82--98}.
\newblock
\begin{APACrefDOI} \doi{10.1017/S0003055417000466} \end{APACrefDOI}
\PrintBackRefs{\CurrentBib}

\bibitem [\protect \citeauthoryear {%
Binmore%
}{%
Binmore%
}{%
{\protect \APACyear {2005}}%
}]{%
binmore2005natural}
\APACinsertmetastar {%
binmore2005natural}%
\begin{APACrefauthors}%
Binmore, K.%
\end{APACrefauthors}%
\unskip\
\newblock
\APACrefYear{2005}.
\newblock
\APACrefbtitle {Natural justice} {Natural justice}.
\newblock
\APACaddressPublisher{}{Oxford university press}.
\PrintBackRefs{\CurrentBib}

\bibitem [\protect \citeauthoryear {%
Bowles%
}{%
Bowles%
}{%
{\protect \APACyear {1998}}%
}]{%
bowles1998endogenous}
\APACinsertmetastar {%
bowles1998endogenous}%
\begin{APACrefauthors}%
Bowles, S.%
\end{APACrefauthors}%
\unskip\
\newblock
\APACrefYearMonthDay{1998}{}{}.
\newblock
{\BBOQ}\APACrefatitle {Endogenous preferences: The cultural consequences of markets and other economic institutions} {Endogenous preferences: The cultural consequences of markets and other economic institutions}.{\BBCQ}
\newblock
\APACjournalVolNumPages{Journal of economic literature}{36}{1}{75--111}.
\PrintBackRefs{\CurrentBib}

\bibitem [\protect \citeauthoryear {%
Boyd%
\ \BBA {} Richerson%
}{%
Boyd%
\ \BBA {} Richerson%
}{%
{\protect \APACyear {2009}}%
}]{%
Boyd2009}
\APACinsertmetastar {%
Boyd2009}%
\begin{APACrefauthors}%
Boyd, R.%
\BCBT {}\ \BBA {} Richerson, P\BPBI J.%
\end{APACrefauthors}%
\unskip\
\newblock
\APACrefYearMonthDay{2009}{}{}.
\newblock
{\BBOQ}\APACrefatitle {Voting with your feet: Payoff biased migration and the evolution of group beneficial behavior} {Voting with your feet: Payoff biased migration and the evolution of group beneficial behavior}.{\BBCQ}
\newblock
\APACjournalVolNumPages{Journal of Theoretical Biology}{257}{2}{331--339}.
\PrintBackRefs{\CurrentBib}

\bibitem [\protect \citeauthoryear {%
Boyd%
\ \BBA {} Richerson%
}{%
Boyd%
\ \BBA {} Richerson%
}{%
{\protect \APACyear {2010}}%
}]{%
boyd2010transmission}
\APACinsertmetastar {%
boyd2010transmission}%
\begin{APACrefauthors}%
Boyd, R.%
\BCBT {}\ \BBA {} Richerson, P\BPBI J.%
\end{APACrefauthors}%
\unskip\
\newblock
\APACrefYearMonthDay{2010}{}{}.
\newblock
{\BBOQ}\APACrefatitle {Transmission coupling mechanisms: cultural group selection} {Transmission coupling mechanisms: cultural group selection}.{\BBCQ}
\newblock
\APACjournalVolNumPages{Philosophical Transactions of the Royal Society B: Biological Sciences}{365}{1559}{3787--3795}.
\PrintBackRefs{\CurrentBib}

\bibitem [\protect \citeauthoryear {%
Djankov%
, Glaeser%
, La~Porta%
, Lopez-de Silanes%
\BCBL {}\ \BBA {} Shleifer%
}{%
Djankov%
\ \protect \BOthers {.}}{%
{\protect \APACyear {2003}}%
}]{%
djankov2003new}
\APACinsertmetastar {%
djankov2003new}%
\begin{APACrefauthors}%
Djankov, S.%
, Glaeser, E.%
, La~Porta, R.%
, Lopez-de Silanes, F.%
\BCBL {}\ \BBA {} Shleifer, A.%
\end{APACrefauthors}%
\unskip\
\newblock
\APACrefYearMonthDay{2003}{}{}.
\newblock
{\BBOQ}\APACrefatitle {The new comparative economics} {The new comparative economics}.{\BBCQ}
\newblock
\APACjournalVolNumPages{Journal of comparative economics}{31}{4}{595--619}.
\PrintBackRefs{\CurrentBib}

\bibitem [\protect \citeauthoryear {%
Fouka%
}{%
Fouka%
}{%
{\protect \APACyear {2020}}%
}]{%
fouka2020backlash}
\APACinsertmetastar {%
fouka2020backlash}%
\begin{APACrefauthors}%
Fouka, V.%
\end{APACrefauthors}%
\unskip\
\newblock
\APACrefYearMonthDay{2020}{}{}.
\newblock
{\BBOQ}\APACrefatitle {Backlash: The unintended effects of language prohibition in US schools after World War I} {Backlash: The unintended effects of language prohibition in us schools after world war i}.{\BBCQ}
\newblock
\APACjournalVolNumPages{The Review of Economic Studies}{87}{1}{204--239}.
\PrintBackRefs{\CurrentBib}

\bibitem [\protect \citeauthoryear {%
Frey%
\ \BBA {} Atkisson%
}{%
Frey%
\ \BBA {} Atkisson%
}{%
{\protect \APACyear {2020}}%
}]{%
Frey2020}
\APACinsertmetastar {%
Frey2020}%
\begin{APACrefauthors}%
Frey, S.%
\BCBT {}\ \BBA {} Atkisson, C.%
\end{APACrefauthors}%
\unskip\
\newblock
\APACrefYearMonthDay{2020}{}{}.
\newblock
{\BBOQ}\APACrefatitle {A dynamic over games drives selfish agents to win–win outcomes} {A dynamic over games drives selfish agents to win–win outcomes}.{\BBCQ}
\newblock
\APACjournalVolNumPages{Proceedings of the Royal Society B}{287}{1941}{20202630}.
\PrintBackRefs{\CurrentBib}

\bibitem [\protect \citeauthoryear {%
Harrison%
\ \protect \BOthers {.}}{%
Harrison%
\ \protect \BOthers {.}}{%
{\protect \APACyear {2020}}%
}]{%
harrison2020gibbs}
\APACinsertmetastar {%
harrison2020gibbs}%
\begin{APACrefauthors}%
Harrison, P.%
, Marjieh, R.%
, Adolfi, F.%
, van Rijn, P.%
, Anglada-Tort, M.%
, Tchernichovski, O.%
\BDBL {}Jacoby, N.%
\end{APACrefauthors}%
\unskip\
\newblock
\APACrefYearMonthDay{2020}{}{}.
\newblock
{\BBOQ}\APACrefatitle {Gibbs sampling with people} {Gibbs sampling with people}.{\BBCQ}
\newblock
\APACjournalVolNumPages{Advances in neural information processing systems}{33}{}{10659--10671}.
\PrintBackRefs{\CurrentBib}

\bibitem [\protect \citeauthoryear {%
Henrich%
}{%
Henrich%
}{%
{\protect \APACyear {2004}}%
}]{%
henrich2004cultural}
\APACinsertmetastar {%
henrich2004cultural}%
\begin{APACrefauthors}%
Henrich, J.%
\end{APACrefauthors}%
\unskip\
\newblock
\APACrefYearMonthDay{2004}{}{}.
\newblock
{\BBOQ}\APACrefatitle {Cultural group selection, coevolutionary processes and large-scale cooperation} {Cultural group selection, coevolutionary processes and large-scale cooperation}.{\BBCQ}
\newblock
\APACjournalVolNumPages{Journal of Economic Behavior \& Organization}{53}{1}{3--35}.
\PrintBackRefs{\CurrentBib}

\bibitem [\protect \citeauthoryear {%
Jancsary%
, Meyer%
, H{\"o}llerer%
\BCBL {}\ \BBA {} Barberio%
}{%
Jancsary%
\ \protect \BOthers {.}}{%
{\protect \APACyear {2017}}%
}]{%
jancsary2017toward}
\APACinsertmetastar {%
jancsary2017toward}%
\begin{APACrefauthors}%
Jancsary, D.%
, Meyer, R\BPBI E.%
, H{\"o}llerer, M\BPBI A.%
\BCBL {}\ \BBA {} Barberio, V.%
\end{APACrefauthors}%
\unskip\
\newblock
\APACrefYearMonthDay{2017}{}{}.
\newblock
{\BBOQ}\APACrefatitle {Toward a structural model of organizational-level institutional pluralism and logic interconnectedness} {Toward a structural model of organizational-level institutional pluralism and logic interconnectedness}.{\BBCQ}
\newblock
\APACjournalVolNumPages{Organization Science}{28}{6}{1150--1167}.
\PrintBackRefs{\CurrentBib}

\bibitem [\protect \citeauthoryear {%
Mitchell%
\ \BBA {} Tetlock%
}{%
Mitchell%
\ \BBA {} Tetlock%
}{%
{\protect \APACyear {2009}}%
}]{%
mitchell2009disentangling}
\APACinsertmetastar {%
mitchell2009disentangling}%
\begin{APACrefauthors}%
Mitchell, G.%
\BCBT {}\ \BBA {} Tetlock, P\BPBI E.%
\end{APACrefauthors}%
\unskip\
\newblock
\APACrefYearMonthDay{2009}{}{}.
\newblock
{\BBOQ}\APACrefatitle {Disentangling reasons and rationalizations: Exploring perceived fairness in hypothetical societies} {Disentangling reasons and rationalizations: Exploring perceived fairness in hypothetical societies}.{\BBCQ}
\newblock
\APACjournalVolNumPages{Social and psychological bases of ideology and system justification}{1}{}{126--158}.
\PrintBackRefs{\CurrentBib}

\bibitem [\protect \citeauthoryear {%
North%
}{%
North%
}{%
{\protect \APACyear {1991}}%
}]{%
North1991}
\APACinsertmetastar {%
North1991}%
\begin{APACrefauthors}%
North, D\BPBI C.%
\end{APACrefauthors}%
\unskip\
\newblock
\APACrefYearMonthDay{1991}{}{}.
\newblock
{\BBOQ}\APACrefatitle {Institutions} {Institutions}.{\BBCQ}
\newblock
\APACjournalVolNumPages{Journal of Economic Perspectives}{5}{1}{97--112}.
\PrintBackRefs{\CurrentBib}

\bibitem [\protect \citeauthoryear {%
Ostrom%
}{%
Ostrom%
}{%
{\protect \APACyear {2005}}%
}]{%
Ostrom2005}
\APACinsertmetastar {%
Ostrom2005}%
\begin{APACrefauthors}%
Ostrom, E.%
\end{APACrefauthors}%
\unskip\
\newblock
\APACrefYear{2005}.
\newblock
\APACrefbtitle {Understanding institutional diversity} {Understanding institutional diversity}.
\newblock
\APACaddressPublisher{}{Princeton University Press}.
\PrintBackRefs{\CurrentBib}

\bibitem [\protect \citeauthoryear {%
Rawls%
}{%
Rawls%
}{%
{\protect \APACyear {1971}}%
}]{%
rawls1971theory}
\APACinsertmetastar {%
rawls1971theory}%
\begin{APACrefauthors}%
Rawls, J.%
\end{APACrefauthors}%
\unskip\
\newblock
\APACrefYear{1971}.
\newblock
\APACrefbtitle {A Theory of Justice} {A theory of justice}.
\newblock
\APACaddressPublisher{Cambridge, Massachusetts}{The Belknap Press of Harvard University Press}.
\PrintBackRefs{\CurrentBib}

\bibitem [\protect \citeauthoryear {%
Richerson%
\ \protect \BOthers {.}}{%
Richerson%
\ \protect \BOthers {.}}{%
{\protect \APACyear {2016}}%
}]{%
Richerson2016}
\APACinsertmetastar {%
Richerson2016}%
\begin{APACrefauthors}%
Richerson, P.%
, Baldini, R.%
, Bell, A\BPBI V.%
, Demps, K.%
, Frost, K.%
, Hillis, V.%
\BDBL {}Newson, L.%
\end{APACrefauthors}%
\unskip\
\newblock
\APACrefYearMonthDay{2016}{}{}.
\newblock
{\BBOQ}\APACrefatitle {Cultural group selection plays an essential role in explaining human cooperation: A sketch of the evidence} {Cultural group selection plays an essential role in explaining human cooperation: A sketch of the evidence}.{\BBCQ}
\newblock
\APACjournalVolNumPages{Behavioral and Brain Sciences}{39}{}{e30}.
\newblock
\begin{APACrefDOI} \doi{10/f3r4wb} \end{APACrefDOI}
\PrintBackRefs{\CurrentBib}

\bibitem [\protect \citeauthoryear {%
Shivakoti%
\ \BBA {} Ostrom%
}{%
Shivakoti%
\ \BBA {} Ostrom%
}{%
{\protect \APACyear {1993}}%
}]{%
Shivakoti1993}
\APACinsertmetastar {%
Shivakoti1993}%
\begin{APACrefauthors}%
Shivakoti, G\BPBI P.%
\BCBT {}\ \BBA {} Ostrom, E.%
\end{APACrefauthors}%
\unskip\
\newblock
\APACrefYearMonthDay{1993}{}{}.
\newblock
{\BBOQ}\APACrefatitle {Farmer and Government Organized Irrigation Systems in Nepal: Preliminary Findings from Analysis of 127 Systems} {Farmer and government organized irrigation systems in nepal: Preliminary findings from analysis of 127 systems}.{\BBCQ}
\newblock
\BIn{} \APACrefbtitle {Common Property in Ecosystems Under Stress, the Fourth Annual Conference of the International Association for the Study of Common Property.} {Common property in ecosystems under stress, the fourth annual conference of the international association for the study of common property.}
\newblock
\APACaddressPublisher{Manila, Philippines}{}.
\newblock
\begin{APACrefURL} \url{http://hdl.handle.net/10535/4957} \end{APACrefURL}
\PrintBackRefs{\CurrentBib}

\bibitem [\protect \citeauthoryear {%
Tetlock%
}{%
Tetlock%
}{%
{\protect \APACyear {1986}}%
}]{%
tetlock1986value}
\APACinsertmetastar {%
tetlock1986value}%
\begin{APACrefauthors}%
Tetlock, P\BPBI E.%
\end{APACrefauthors}%
\unskip\
\newblock
\APACrefYearMonthDay{1986}{}{}.
\newblock
{\BBOQ}\APACrefatitle {A value pluralism model of ideological reasoning.} {A value pluralism model of ideological reasoning.}{\BBCQ}
\newblock
\APACjournalVolNumPages{Journal of personality and social psychology}{50}{4}{819}.
\PrintBackRefs{\CurrentBib}

\bibitem [\protect \citeauthoryear {%
Wildavsky%
}{%
Wildavsky%
}{%
{\protect \APACyear {1987}}%
}]{%
wildavsky1987choosing}
\APACinsertmetastar {%
wildavsky1987choosing}%
\begin{APACrefauthors}%
Wildavsky, A.%
\end{APACrefauthors}%
\unskip\
\newblock
\APACrefYearMonthDay{1987}{}{}.
\newblock
{\BBOQ}\APACrefatitle {Choosing preferences by constructing institutions: A cultural theory of preference formation} {Choosing preferences by constructing institutions: A cultural theory of preference formation}.{\BBCQ}
\newblock
\APACjournalVolNumPages{American political science review}{81}{1}{3--21}.
\PrintBackRefs{\CurrentBib}

\bibitem [\protect \citeauthoryear {%
Zhong%
\ \BBA {} Frey%
}{%
Zhong%
\ \BBA {} Frey%
}{%
{\protect \APACyear {2022}}%
}]{%
zhong2022institutional}
\APACinsertmetastar {%
zhong2022institutional}%
\begin{APACrefauthors}%
Zhong, Q.%
\BCBT {}\ \BBA {} Frey, S.%
\end{APACrefauthors}%
\unskip\
\newblock
\APACrefYearMonthDay{2022}{}{}.
\newblock
{\BBOQ}\APACrefatitle {Institutional similarity drives cultural similarity among online communities} {Institutional similarity drives cultural similarity among online communities}.{\BBCQ}
\newblock
\APACjournalVolNumPages{Scientific reports}{12}{1}{18982}.
\PrintBackRefs{\CurrentBib}

\end{thebibliography}

\end{document}